\documentclass{article}

\usepackage{graphicx}
\usepackage{bm}
\usepackage{amssymb}
\usepackage{amsmath}
\usepackage{cite}
\usepackage{enumitem}

\author{Aron C. Wall\footnote{aroncwall@gmail.com}
\\ \textit{Department of Physics} \\ \textit{University of California, Santa Barbara}
\\ \textit{Santa Barbara, CA 93106, USA} }
\title{Maximin Surfaces, and the Strong Subadditivity of the Covariant Holographic Entanglement Entropy}

\date{\today}

\begin{document}

\maketitle

\begin{abstract}
The covariant holographic entropy conjecture of AdS/CFT relates the entropy of a boundary region $R$ to the area of an extremal surface in the bulk spacetime.  This extremal surface can be obtained by a maximin construction, allowing many new results to be proven.  On manifolds obeying the null curvature condition, these extremal surfaces: i) always lie outside the causal wedge of $R$, ii) have less area than the bifurcation surface of the causal wedge, iii) move away from the boundary as $R$ grows, and iv) obey strong subadditivity and monogamy of mutual information.  These results suggest that the information in $R$ allows the bulk to be reconstructed all the way up to the extremal area surface.  The maximin surfaces are shown to exist on spacetimes without horizons, and on black hole spacetimes with Kasner-like singularities.
\end{abstract}

\vspace{-.5cm}

\tableofcontents

\section{Introduction}\label{intro}

AdS/CFT is a conjectured duality relating an asymptotically AdS quantum gravity theory (the ``bulk'') to a conformal field theory living on the AdS-boundary of the spacetime.  In its best established form \cite{AdSCFT}, the CFT is a super-Yang-Mills theory with a large number $N$ of colors.  In this case, certain thermodynamic states correspond to \emph{classical} bulk manifolds, described by general relativity, coupled to certain matter fields.  In this limit, the matter fields obey the null energy condition $T_{ab} k^a k^b \ge 0$, where $k^a$ is a null vector.  Thus, one learns interesting facts about large $N$ quantum field theory from the application of general relativity results.

In cases where the bulk spacetime contains an eternal black hole, the boundary consists of two CFT's in an entangled state.  Restricting to a single CFT, one obtains a thermal state \cite{maldacena01}, whose entropy is given by the Bekenstein-Hawking entropy
\begin{equation}\label{BH}
S_\mathrm{CFT} = S_\mathrm{BH} = \frac{ \mathrm{Area} [ \mathrm{Horizon} ] }{4\hbar G}.
\end{equation}

\paragraph{Static Holographic Entropy} The ``holographic entanglement entropy'' conjecture of Ryu and Takayanagi \cite{RT06} generalizes this entropy-area relation to the case of particular regions of the CFT.  It applies only to static manifolds (which can be foliated into static time slices in a canonical way).  It says that on any static time slice, the entropy $S_A$ of a region $A$ in the CFT is proportional to the area of a minimal area surface $\mathrm{min}(A)$ anchored to $A$:
\begin{equation}\label{minA}
S_{A} = \frac{\mathrm{Area}[\mathrm{min}(A)]}{4\hbar G}.
\end{equation}
Technically, both sides of this equation are infinite: the left-hand side because of the divergence of entanglement entropy near a sharp boundary in QFT \cite{diverge}, the right-hand side because of the fact that the surface extends to the AdS boundary which is at an infinite distance.  However, one can get good results by regulating both divergences together and then comparing universal logarithmic or finite contributions \cite{RT06}.  Note also that in the case where $A$ is the whole spacetime, the situation reduces to Eq. (\ref{BH}).  

\begin{figure}[ht]
\centering
\includegraphics[width=.5\textwidth]{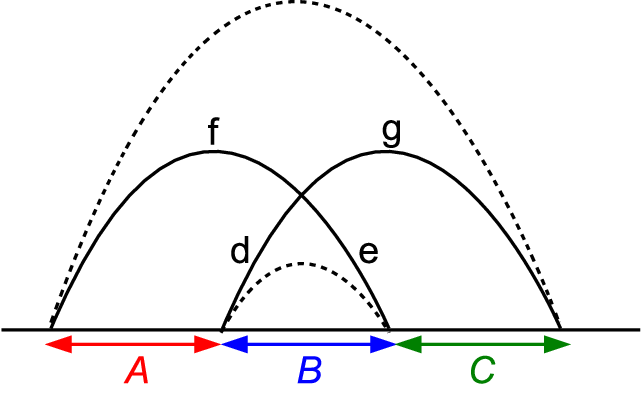}
\caption{\footnotesize The picture proof of Strong Subadditivity for the Ryu-Takayanagi conjecture on a static slice \cite{HT07}.  The horizontal line represents the boundary, the solid lines are the surfaces $\mathrm{min}(AB)$ and $\mathrm{min}(BC)$, and the dashed lines are the surfaces $\mathrm{min}(ABC)$ and $\mathrm{min}(B)$.  The area of $\mathrm{min}(B)$ is less than $de$, while the area of $\mathrm{min}(ABC)$ is less than $fg$.  (Note that the difference between these areas cannot diverge at the boundary; otherwise it would be more area-efficient for $\mathrm{min}(AB)$ and $\mathrm{min}(BC)$ to coincide exactly with $\mathrm{min}(ABC)$ and $\mathrm{min}(B)$ in a neighborhood of the boundary.)}\label{SSAfig}
\end{figure}

In order for relation (\ref{minA}) to be true, the right-hand side of the equation has to satisfy the quantum information condition known as Strong Subadditivity \cite{LR73}, which says that for any three disjoint regions $A$, $B$, and $C$,
\begin{equation}
S_{AB} + S_{BC} \ge S_{ABC} + S_{B}.
\end{equation}
In quantum field theory, each of these entropies are divergent, but the divergences cancel on the left- and right-hand sides of the equation since both sides share the same boundaries---and the same holds for the areas in the bulk dual.  The corresponding Strong Subadditivity relation for $\mathrm{Area}[\mathrm{min}]$ was proven by Ref. \cite{HT07} (some specific examples were shown in Ref. \cite{HiT07}).  The picture proof follows almost trivially from Figure \ref{SSAfig}.  However, it relies critically on the facts that i) $\mathrm{min}[ABC]$ and $\mathrm{min}[B]$ are minimal surfaces ii) lying on the \emph{same} time slice as $\mathrm{min}[AB]$ and $\mathrm{min}[BC]$.  These facts do not apply to the covariant generalization of the Ryu-Takayanagi conjecture, described next.
\paragraph{Covariant Holographic Entanglement Entropy} The original version of the Ryu-Takayanagi conjecture only applies at one moment of time on a static manifold.  This makes it inapplicable on dynamically evolving spacetimes, or to choices of $A$ which do not correspond to static time-slices \cite{CHH12}.  The problem is that there are no slices which are minimal in time, since wiggling in the time direction typically decreases the area.  So Hubeny, Rangamani, and Takayanagi (HRT) \cite{HRT07} suggested a generalization: instead of looking for the minimal area surface, look for the extremal area surface $m(A)$ (if there is more than one, choose the surface with the least area).   

In this article, Strong Subadditivity will be proven for the HRT conjecture (Theorem \ref{SSA}), assuming the null curvature condition $R_{kk} \ge 0$ (NCC).  (On spacetimes violating NCC, the covariant version of the conjecture does not always hold \cite{AT11,CHH12}, despite the fact that the static proof can be proven without using the NCC)  This result is consistent with investigations of Strong Subadditivity in 2+1 dimensional AdS-Vaidya spacetimes \cite{AT11,CHH12}.

Using the same method, the tripartite relation known as the ``monogamy of the mutual information'' can also be proven for HRT surfaces (Theorem \ref{mono}).  This states that for any three disjoint regions $A$, $B$, and $C$,
\begin{align}
\mathrm{Area}[m(AB)] + \mathrm{Area}[m(BC)] + \mathrm{Area}[m(AC)] \ge \phantom{Mn} \notag \\ 
\mathrm{Area}[m(A)] + \mathrm{Area}[m(B)] + \mathrm{Area}[m(C)]  + \mathrm{Area}[m(ABC)].
\end{align}
This result was shown for the static Ryu-Takayanagi surfaces by Ref. \cite{HHM11}.  Evidence for its validity in the covariant case was presented by Refs. \cite{AT11,BBCCG11}.  This relation is \emph{not} valid for general quantum subsystems, but nevertheless holds for the holographic entanglement entropy.

\paragraph{Duality of Holographic Observables}
Given a region $A$ on the boundary, what region in the bulk can be reconstructed?  It has been argued that this region must include at least the causal wedge $I^-(D_A) \cap I^+(D_A)$ (i.e. the intersection of the past and future of the domain of dependence of $A$) \cite{HR12, CKNR12, BLR12}.  This is the region that could be seen by an observer who starts and ends in $D_A$.  However, it is possible $A$ might allow one to reconstruct a larger region.  Following Ref. \cite{CKNR12}, I propose that one can fully reconstruct the spacetime region lying spatially in between $m(A)$ and $D_A$---I will call this region $r(A)$.\footnote{The proposal of Ref. \cite{CKNR12} is actually slightly weaker than this, namely that one can reconstruct only the region spanned by all surfaces $m(B)$ with $B \subset A$.  This is not exactly the same, since in certain cases $m(A)$ can jump discontinuously as the region $A$ continuously grows.  This might leave ``holes'' inside the region $r(A)$.  Ref. \cite{CKNR12} observed that in certain cases one could also reconstruct these holes.  Thus it seems simplest to propose that the entire region $r(A)$ can be reconstructed.  Note also that when $A$ is chosen at the discontinuous phase transition, there are two extremal surfaces with the exact same area.  In this case, I express no opinion as to which of the extremal surfaces should be used.}

Some reasons for making this Dual Observables conjecture are as follows:  While the entropy $S_A = -\mathrm{tr}(\rho_A\,\ln\,\rho_A)$ does not by itself determine the full state $\rho_A$, nevertheless $S_A$ is sensitive to every kind of degree of freedom that is in $A$, and does not depend on any of the degrees of freedom outside of $A$.   If there were any degrees of freedom in $r(A)$ besides those in $A$, then it seems odd that value of $\mathrm{Area}[m(A)]$ should be insensitive to those degrees of freedom.  

Furthermore, it is highly suggestive that the leading order divergence in the entanglement entropy of a region is proportional to the area of its boundary \cite{diverge}.  If quantum gravity cuts off this divergence at the Planck scale, this could naturally lead to a finite entropy per area, not only for slices of black hole horizons \cite{area}, but also for general codimension 2 surfaces.\footnote{However, if one assigns a Bekenstein-Hawking entropy to all codimension 2 surfaces, one should note the puzzling fact that the Second Law is violated on codimension 1 null surfaces \cite{10proofs}, unless the null surface is a causal horizon \cite{JP03}.  Classically, an extremal surface such as $m(A)$ represents the time at which the entropy switches from increasing to decreasing.  Could this be why extremal surfaces are special?} Thus, if it were true that the observables in $r(A)$ were dual to $D_A$, it could naturally explain the reason for the success of the HRT conjecture.

However, the conjectured duality of $D_A$ to $r(A)$ has some nontrivial consequences for the extremal area surfaces $m(A)$.  For example, $m(A)$ must always lie spatially outside of the causal wedge $I^-(D_A) \cap I^+(D_A)$ (otherwise it would be possible for signals to causally propagate from outside of $r(A)$ into $D_A$, in violation of the Dual Observables conjecture).  That $m(A)$ is spacelike outside will be proven in Theorem \ref{wm} (see also Ref. \cite{HR12}).

A second consistency condition is that as the region $A$ on the boundary grows, $r(A)$ must move spatially outwards, away from the boundary region $A$.\footnote{I shall use ``outwards'' to refer to the direction going away from $A$ towards the complementary boundary region $\overline{A}$, and ``inwards'' to mean the opposite direction.  This may be counterintuitive to those who expect that the boundary $A$ should be outside the bulk, but either convention requires labelling some region on the boundary as the ``inside''.  I have picked this convention to match the notion of being ``inside'' or ``outside'' the causal wedge.}  Otherwise the field observables measurable in $r(A)$ would not monotonically increase, violating the conjecture.  This outward-monotonicity condition was conjectured by Ref. \cite{CKNR12} and will be proven here as Theorem \ref{outward}.  The fact that the Dual Observables conjecture passes all of these consistency conditions provides some evidence that it is true.

On the other hand, the Dual Observables conjecture has strange consequences when applied to a ``bag of gold'' spacetime, consisting of a closed FRW universe connected by a (nontraversable) wormhole to an AdS-Schwarzschild geometry.  Semiclassically, one can store an arbitrarily large amount of entropy in the FRW universe \cite{sorkin97,jacobson99,marolf08,FHMMRS05}, which cannot be accounted for by the CFT, suggesting that some observables in the bulk are superselection sectors \cite{MW12}.  This picture would work very nicely if $m(\mathrm{CFT})$ were the extremal surface lying in the throat of the wormhole.  But actually there is an extremal area surface with even less entropy: the empty set!  So maybe the covariant conjecture needs to be adjusted for spacetimes with global horizons.  But in this article, we will not consider such modifications.


\paragraph{Methodology of the Proofs} The main idea of the proof is to rephrase the covariant version of the Ryu-Takayanagi conjecture in a different way: Instead of focusing on the extremal HRT surface $m(A)$, let us focus on the maximin surface $M(A)$, which is defined by \emph{minimizing} the area on some achronal slice $\Sigma$, and then \emph{maximizing} the area with respect to varying $\Sigma$.  The surface $M(A)$ is more convenient for proving inequalities such as Strong Subadditivity, since like the static Ryu-Takayanagi surface it is defined as a minimum on some slice.  Using the power of the extreme value theorem, it can even be shown to exist (Theorems \ref{Mexists}-\ref{Kas}) for certain broad classes of spacetimes.  However, Theorem \ref{Mism} shows that $M(A) = m(A)$, i.e. the two definitions are equivalent (for spacetimes obeying the NCC).

While the level of rigor of the following proofs is elevated compared to the usual standard in physics, it is not intended to be completely watertight according to the standard of mathematics.  Since the procedure to define $M$ involves varying over an infinite-dimensional space of all possible surfaces $s$ and achronal slices $\Sigma$, I have tried to make the existence proofs valid even when $s$ and $\Sigma$ are horribly wiggly (e.g. if they have no tangent plane).  However, I have been more lax when dealing with quantities such as the expansion $\theta$ (or the extrinsic curvature $K$).  Null surfaces tend to form cusps at which $\theta$ is ill-defined, but I assume that this could be dealt with by defining $\theta = \pm \infty$ as appropriate, and treating them by analogy to the finite case.  I have also been cavalier with the renormalization procedure used to define quantities such as $\mathrm{Area}[M]$, which will be manipulated as if finite.  

\section{Assumptions about the Spacetime}

The bulk spacetime will be assumed to be classical, smooth, and asymptotically locally AdS.\footnote{H. Casini has pointed out to me that Strong Subadditivity is violated for extremal surfaces in a spacetime whose boundary is at finite distance, e.g. a cylinder cut out of Minkowski spacetime.  The first step to fail is 6(c), because the Second Law is valid only for causal horizons, i.e. the boundary of the past of points at infinity.}  It will be required to obey the NCC $R_{ab} k^a k^b \ge 0$ for any null vector $k^a$.  Except where explicitly noted, a generic condition will also be assumed, namely that there exists nonvanishing null-curvature $R_{ab} k^a k^b$ or shear $\sigma_{ab}$ along at least one point of any segment of any null ray (lying on some null surface $N$).  For most physical questions of interest, continuity implies that the generic condition can be dropped, so long as one replaces certain strict inequalities with weak inequalities.

The spacetime will also be assumed to be AdS-hyperbolic.  By this I mean that i) there are no closed causal curves, and ii) for any two points $x$ and $y$, $\mathcal{I}^+(x) \cap \mathcal{I}^-(y)$ is compact after conformally compactifying the AdS boundary.  This condition parallels the definition of global hyperbolicity \cite{BS07}, except at the AdS boundary, which is not globally hyperbolic by the usual definition.  This allows the spacetime to be foliated by a time function $t$ \cite{geroch67}.  It will also be assumed that space at one time is compact, after compactifying the AdS boundary.

Except in Theorem $\ref{Mexists}$, we allow the spacetime to include black holes, although some constraints on their singularity structure are needed in Theorem \ref{Kas}.\footnote{In a previous version of this article, I said that the results only applied to horizonless spacetimes.  That was because I could only prove the existence of maximin surfaces when there are no horizons.  However, I now believe that this was a tactical mistake, because it caused many people to think that the proof of strong subadditivity, and the other results, only apply to such spacetimes.  However, so long as one is willing to assume the existence of maximin surfaces, all of the proofs still apply even to black hole spacetimes.  (For comparison, most of the previous articles proving general results about holographic entanglement entropy do not bother to prove the general existence of the surfaces in question.)  Furthermore, the new Theorem \ref{Kas} now proves the existence of maximin surfaces for certain black hole metrics such as eternal AdS-Schwarzschild.} All codimensions are defined relative to the bulk spacetime dimension.


\section{Results}

In order to keep this section self-contained, the main concepts used in the proofs below will be defined again, with greater precision.  Some necessary definitions and lemmas are in section \ref{pre}.  An key concept is the idea that any extremal surface $x$ has a ``representative'' $\tilde{x}$ defined on any other time slice $\Sigma$, defined by shooting out a null surface from $x$.  This allows one to project all extremal surfaces onto the same slice $\Sigma$, enabling one to construct proofs in the dynamical case which are analogous to the static proofs.

It will then be shown that the surface is farther from the boundary than the causal wedge, and is spacelike separated from it (section \ref{causal}).  In addition to being interesting in its own right, this theorem plays an important role in being able to establish that the aforesaid representatives $\tilde{x}$ actually exist.

Section \ref{def} describes the maximin construction: the core idea of this article.

Section \ref{exist} shows that in certain classes of spacetimes there actually exist maximin surfaces.  This includes not only horizonless spacetimes, but also singularities governed by the Kasner metric (as well as other homogeneous power-law cosmologies which satisfy the null curvature condition).  Section \ref{stable} gives conditions for a technical stability property needed in the proofs that follow.  Readers who are only interested in the main results may wish to skip these two sections.  So long as one is willing to simply \emph{assume} the existence of stable maximin surfaces, these sections are unnecessary.  (By comparison, most work on extremal HRT surfaces simply assumes that the surfaces in question exist.)

Section \ref{equiv} proves the equivalence of the maximin surfaces to the HRT surfaces, and section \ref{props} proves that they have more area than the causal surface, move outwards monotonically under growth of the boundary region, obey strong subadditivity, and monogamy of mutual information.

\subsection{Preliminary Definitions and Lemmas}\label{pre}

\begin{enumerate}

\item \label{1} Definitions: Given a boundary spatial region $A$, the HRT surface $m(A)$ is the codimension 2 surface with extremal area which is anchored to the spatial boundary $\partial_0(D_A) = \partial A$ of the causal domain of dependence $D_A$ for $A$ \cite{HRT07}.  ($D_A$ may be defined as the set of points $x$ on the boundary such that every inextendible timelike worldline passing through $x$ also passes through $D_A$.)  In the case where the spacetime has a nontrivial codimension-2 holomology, we also require $m(A)$ to be homologous to $A$ \cite{HT07,fursaev06}.\footnote{Note that spacetimes with nontrivial codimension-2 homology always have horizons, by the Topological Censorship Theorem \cite{top}.}  Let $r(A)$ be the spacetime region which is spatially in between $m(A)$ and $D_A$ (i.e. r(A) is the region which is spacelike separated to $m(A)$ and on the same side as $A$.)
	\begin{enumerate}
	\item If there are multiple such extremal surfaces $x_1(A),\,x_2(A) \ldots$, then $m(A)$ is the extremal surface with the least area.
	\item If there is a tie for least area, then let all the tied surfaces $m_1(A),\,m_2(A) \ldots$ be considered HRT surfaces.  (Ties which occur in the following definitions will be handled in a similar way).
	\end{enumerate}
\item Definitions: Given any codimension 2 extremal surface $x(A)$ as defined above (not necessarily the one with minimal area), one can shoot out codimension 1 null surfaces $N(A)$ in any of the four null directions (past-inward, past-outward, future-inward, future-outward) from $x(A)$.  (In the case of $m(A)$, if one chooses the ``inward'' direction in which N(A) is shot towards $D_A$ rather than $D_{\bar{A}}$, then $N(A) = \partial r(A)$.)  Although $x(A)$ does not lie on every complete achronal slice $\Sigma$, one can still define a codimension 2 ``representative'' on $\Sigma$ defined as $\tilde{x}(A,\Sigma) = N(A) \cap \Sigma$.
	\begin{enumerate}
	\item We will only be interested in $\tilde{x}(A,\Sigma)$ in the case where $\partial A \in \Sigma$, so that $\tilde{x}(A,\Sigma)$ is anchored to the boundary in the same place as $x(A)$.  	
	\end{enumerate}

\item \label{trap} Theorem: The representative $\tilde{x}(A,\Sigma)$ has less area than $x(A)$ (unless it \emph{is} $x(A)$).  Proof: Because $x(A)$ is extremal, the null surfaces $N(A)$ have expansion $\theta = 0$ at $x(A)$.  From this point on the proof parallels the standard result \cite{HawkingEllis} for trapped surfaces: By the Raychaudhuri equation, the derivative of the expansion is
\begin{equation}\label{Ray}
\frac{d\theta}{d\lambda} = -\frac{\theta^2}{D-2} - \sigma_{ab}\sigma^{ab} - R_{ab} k^a k^b,
\end{equation}
where $\lambda$ is an affine parameter, $k^a$ is a null vector which is unit with respect to $\lambda$, and $\sigma_{ab}$ is the shear part of the null extrinsic curvature.  By the null curvature condition together with the generic condition, the null rays must focus, so that $\theta < 0$ everywhere on $N(A)$ (except at $x(A)$ itself).  Hence the area of any set of null generators in $N(A)$ decreases when moving away from $x(A)$.  When null generators intersect (as a result of caustics where $\theta \to -\infty$, or when distant null generators collide) they can exit $N(A)$, but no new generators can enter.  Hence the total area is decreasing, and any slice $\tilde{x}(A,\Sigma)$ of $N(A)$ has less area than $x(A)$.

\item \label{thetaineq} Theorem: Let $N_1$ and $N_2$ be null congruences.  Let $N_2$ be nowhere to the past of $N_1$
(i.e. $N_2 \cap I^-(N_1)$ is empty), and let them touch at the point $x$ on a slice $\Sigma$.  Then in any sufficiently small neighborhood of $x$, either i) $N_1$ and $N_2$ coincide or ii) there exists a point $y$ for which $N_2$ is expanding faster, i.e. $\theta[N_2] > \theta[N_1]$ (cf. Fig. \ref{bend}).  The proof below follows Theorem 1 of Ref. \cite{sing} (which proves a similar monotonicity result for the generalized entropy).
\begin{figure}[hbt]
\centering
\includegraphics[width=.4\textwidth]{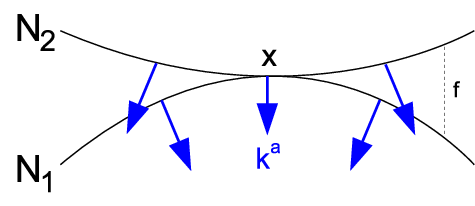}
\caption{\footnotesize Two null surfaces $N_1$ and $N_2$ are pictured as they appear at one time, on the slice $\Sigma$.  $N_1$ is nowhere outside of $N_2$, and coincides with $N_2$ at $x$.  The generating null vectors $k^a$, projected onto the slice $\Sigma$, must be normal to the null surfaces.  $f$ is the proper distance between the two null surfaces $N_1$ and $N_2$, viewed as a function of $N_1$ or $N_2$.  Because $N_2$ can only bend outwards relative to $N_1$ at $x$, it must be expanding faster than $N_1$ in some neighborhood of the point $x$ (unless the surfaces coincide exactly in some neighborhood of $x$).}\label{bend}
\end{figure}
	\begin{enumerate}
	\item Since $N_1$ and $N_2$ coincide and are smooth at a point $x$, and $N_2$ cannot cross $N_1$, $N_1$ and $N_2$ must share the same tangent plane.  The null extrinsic curvature of a null surface is defined as:
\begin{equation}\label{B}
B_{ab} = h^c_a h_{bd} \nabla_c k^d,
\end{equation} 
where $h_{ab}$ is the pullback of the metric tensor onto the co-dimension 2 surface $\Sigma \cap N$, and $k^a$ is a (future-oriented) null vector pointing in the direction of the null generators on $N$.  After subtracting off the contribution of the extrinsic curvature of $\Sigma$ itself (which is the same for both $N_1$ and $N_2$), $B_{ab}$ is proportional to the spatial extrinsic curvature $K_{ab}$ of $\Sigma \cap N$ in $\Sigma$.  For any vector $v_a$ and point $x$, the extrinsic curvature component $K_{ab} v^a v^b (x)$ measures how much the surface $N$ curves away from its tangent plane, to second order, as one travels away from $x$ in the direction of $v^a$.  The expansion of a null surface is related to the null extrinsic curvature as follows:
\begin{equation}\label{exp}
\theta \equiv (\mathrm{Area})^{-1} k^a \nabla_a \mathrm{Area} = B_{ab} h^{ab}.
\end{equation}
	\item On the achronal slice $\Sigma$, let the shortest proper distance between the surfaces $N_1$ and $N_2$ be given by a smooth function $f(N_1) \ge 0$.  Since the tangent planes of $N_1$ and $N_2$ coincide at $x$, $f$ vanishes to zeroth and first order as one moves away from $x$.  Hence, in a neighborhood with lengthscale $\epsilon$, $f \lesssim \epsilon^2 \ll \epsilon$, so the two surfaces are very close.

One can map points on $N_1$ to points on $N_2$ by choosing a set of $D-2$ coordinates on the $D$-dimensional bulk manifold; if these coordinates vary smoothly, then their values on $N_1$ and $N_2$ are ambiguous only up to shifts of order $f$ or less.   Hence, up to subleading corrections the function can be defined on either of the two null surfaces: $f(N_1) = f(N_2) + \mathcal{O}(f \nabla f)$.  Since $\nabla f \lesssim \epsilon$, the ambiguity in $f$ is always subleading compared to $f$ itself.

This mapping of points on $N_1$ and $N_2$ also allows the null generating vectors $k^a$ to be compared on the corresponding points of $N_1$ and $N_2$ (Fig. \ref{bend}).  When the $k^a$ of $N_1$ or $N_2$ is projected onto $\Sigma$, it must be normal to that surface, because a lightfront always travels in the direction perpendicular the front itself.  So $k^a|_\Sigma = c n^a$, where $n^a$ is an inward pointing normal vector and $c > 0$ is an arbitrary real number.  To facilitate comparisons of different $k^a$ vectors, we will choose $c = 1$ everywhere on $N_2$ and $N_1$.
	\item For small $\nabla f$, the difference between $k^a$ on $N_1$ and $N_2$ is given by
\begin{equation}
\Delta k^a = k^{a}[N_2] - k^{a}[N_1] = \nabla^a f + \mathcal{O}((\nabla f)^2).
\end{equation}
This is the linearized approximation to the standard trigonometric identity expressing the components of a rotated unit vector in terms of sines and cosines.  Neglecting the higher order cosine terms, $\Delta k^a$ lies on the codimension 2 surface $N_2 \cap \Sigma$ (or $N_1 \cap \Sigma$).  The extrinsic curvature difference can now be calculated from Eq. (\ref{B}):
\begin{equation}
\Delta B_{ab} = B_{ab}[N_2] - B_{ab}[N_1] = \nabla_a \nabla_b f.
\end{equation}
We then contract with respect to the inverse metric $h^{ab}$ pulled back to the null surface\footnote{To leading order in $\epsilon$ it does not matter whether we use the metric on $N_1$ or $N_2$.}, and use Eq. (\ref{exp}) to find that
\begin{equation}
\Delta \theta = \theta[N_2] - \theta[N_1] = \nabla^2 f,
\end{equation}
a total derivative.  

Let $N_1$ (or $N_2$) be labelled by an $r$ coordinate representing the proper distance from $x$, and let $d\sigma$ be the volume element on the codimension 3 space of constant $r$ on $N_2 \cap \Sigma$.  Let us define a Green's function $G(y)$ on the ball of points $y$ with $r < R$, to be the solution to these equations:
\begin{equation}
-\nabla^2 G(y) = \delta^{D-2}(y); \qquad G|_{r=R} = 0,
\end{equation}
where $\nabla$ is defined on $N_1$ (or $N_2$).  For a sufficiently small $R$, the metric $h^{ab}$ is very close to being a flat Euclidean metric, so that $G \propto (r^{D-4} - R^{D-4})/(D-4)$ (or $\ln(R/r)$ in $D = 4$).  In any dimension, $G(y) > 0$ for $r < R$, and thus $\partial_r G|_{r=R} < 0$.  For sufficiently small $R$ these inequalities must continue to hold if the metric is slightly deformed by nonzero curvature.

One can now use $G$ to integrate $\Delta \theta$ on the codimension 2 ball $B$:
\begin{equation}\label{integral}
\int_B G\,\Delta \theta\,d^{D-2}y = \int_B G\,\nabla^2 f d^{D-2}y
= -\int_{\partial_B} f\,\partial_r G \,d\sigma\ge 0.
\end{equation}
where we have integrated by parts twice and used the fact that $f(0) = 0$.

Now either (i) $f = 0$ in a neighborhood of $x$, or else (ii) for arbitrarily small values of $R$, the right hand side of Eq. (\ref{integral}) is strictly positive, in which case $\Delta \theta$ must also be positive for at least some points arbitrarily close to $x$.  This proves the theorem.

	\item \label{Kineq} Corollary: By the same argument, using $n^a$ in place of $k^a$, either i) $N_1$ and $N_2$ coincide in a neighborhood of $x$ or ii) there exists a point $y$ arbitrarily close to $x$ for which $\mathrm{tr}(K)[N_2] > \mathrm{tr}(K)[N_1]$, where $\mathrm{tr}(K)$ is the trace of the extrinsic curvature of $N_1$ or $N_2$, restricted to the slice $\Sigma$.
	\item \label{elliptical} Corollary: If in addition, $\theta[N_1] \ge 0$ and $\theta[N_2] \le 0$, then the two surfaces must coincide in every part that is connected to the point $x$.  The same is true using $\mathrm{tr}(K)$ in place of $\theta$.
	\end{enumerate}

\end{enumerate}
\subsection{Extremal Surfaces lie outside Causal Surfaces}\label{causal}

For any region $D_A$ in the boundary CFT, there is a natural causal domain of dependence associated with bulk causality.  The edge of this causal wedge is called the causal surface.

In this section, we will show that extremal surfaces lie farther away from the boundary than the causal surface does.  This proof uses the NCC.  We will show that the extremal surface is in fact \emph{spacelike} outside the causal surface.  (The weaker statement that $x(A)$ does not lie in the interior of the causal wedge $I^-(D_A) \cap I^+(D_A)$ was independently shown in Ref. \cite{HR12}.)

This suggests that likely one can use the CFT to reconstruct a bigger region than just the causal wedge, as discussed in section \ref{intro}.

\begin{enumerate}[resume]
\item Definition: Let the causal surface $w(A)$ be defined as the intersection between the past and future horizons of the domain of dependence of $A$: $w(A) = \partial I^-(D_A) \cap \partial I^+(D_A)$.  This surface is a codimension 2 spacelike surface lying at the edge of the causal wedge $I^-(D_A) \cap I^+(D_A)$ \cite{HRT07}.

\item\label{wm} Theorem: An extremal surface $x(A)$ lies outside of $w(A)$, in a spacelike direction. 
	\begin{enumerate}
	\item Since $\partial I^-(D_A)$ is a future causal horizon, by the usual argument \cite{HawkingEllis} it satisfies the classical Second Law, so that the area must increase on slices of $\partial I^-(D_A)$ when moving to the future away from $w(A)$.  The proof is reminiscent of Theorem \ref{trap}: If at any point on the horizon $\theta < 0$, by the Raychaudhuri Eq. (\ref{Ray}) the null generators would have to focus, causing them to eventually meet each other, exiting the horizon.  But on an AdS-hyperbolic spacetime, null generators cannot exit a future horizon, since it is defined by shooting rays back from null infinity \cite{HawkingEllis}.  Null generators can enter the horizon, but this only causes the area to increase even more.  So the area is always increasing.

Similarly, the area increases on slices of the past horizon $\partial I^+(D_A)$ when moving to the past away from $w(A)$.

	\item Suppose that $x(A)$ is not entirely outside $w(A)$ (Fig. \ref{wvsm}).  Let $N(A)$ be the null congruence shot out from $x(A)$ (towards the $A$ side of the boundary).  Consider a one parameter family of boundary spacetime regions $R(q)$ continuously interpolating between $R(1) = D_A$ and a smaller region $R(0)$ lying well within the region $D_A$.\footnote{Note that there is no requirement that $R(q)$ be the domain of dependence $D_B$ for any region $B$.  Otherwise we might have to worry about the fact that domains of dependence change discontinuously in some cases.}  The associated causal surfaces $w(R(q)) = \partial I^-(R(q)) \cap \partial I^+(R(q))$ also change continuously as $q$ is adjusted.  If we choose $R(0)$ to be small enough that $w(R(q))$ lies inside $x(A)$, then by continuity, at some intermediate value $q_*$, $w(R(q_*))$ touches $N(A)$, but is nowhere outside of it.  From this it is possible to derive a contradiction.
\begin{figure}[ht]
\centering
\includegraphics[width=.47\textwidth]{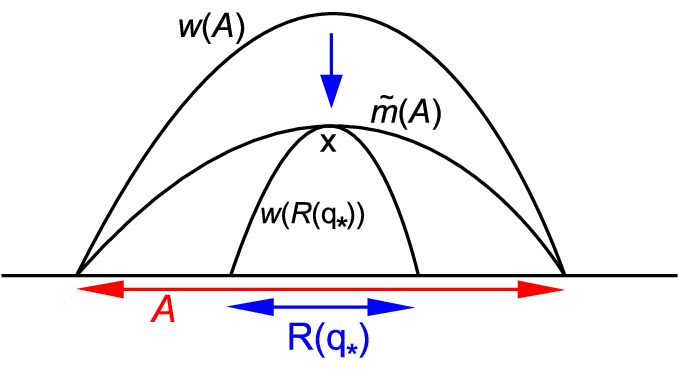}
\caption{\footnotesize The causal surface $w(A)$ is pictured on an achronal slice $\Sigma$.  If $m(A)$ does not lie entirely outside of $w(A)$, then by continuously shrinking $D_A$ one can find a region $R(q_*)$ whose causal wedge touches the representative $\tilde{x}(A)$ on $\Sigma$ at some point $x$, which is contradictory.}\label{wvsm}
\end{figure}\label{wvsm}
	\item $w(R(q_*))$, being a slice of a causal horizon, has $\theta > 0$ by the Second Law.  On the other hand, $N(A)$ has $\theta < 0$ by Theorem \ref{trap}.  (These inequalities are strict due to the generic condition.) Where the two surfaces touch, the fact that $N(A)$ is nowhere inside $w(R(q_*))$ means that the null extrinsic curvature of $N(A)$ must bend outwards at least as much as the null extrinsic curvature of $w(R(q_*))$, by Theorem \ref{thetaineq}.  Since $\theta$ is the trace of the null extrinsic curvature, one has $\theta[N(A)] \ge \theta[w(R(q_*))]$, contradicting the previous statements.
	\item Comment: If one does not assume the generic condition, then $w(A)$ might touch some representative $\tilde{x}(A)$ of $x(A)$.  But if so, by Corollary \ref{elliptical} the connected components that touch must entirely coincide.
	\item \label{repex} Corollary: On any complete achronal slice $\Sigma$ that passes through the anchor points $\partial A$, the representative $\tilde{x}(A,\Sigma)$ exists, because $N(A)$ is outside of $I^-(D_A) \cap I^+(D_A)$.  (By contrast, for sufficiently late slices $\Sigma$ that do not pass through $\partial A$, $N(A)$ may have already hit the AdS boundary and disappeared.)  Also, $\tilde{x}(A,\Sigma)$ is homologous to $x(A)$, because they are connected by $N(A)$.
	\item Corollary: the causal region $w(A)$ and the causal region of its complement $w(\bar{A})$ do not overlap.  This is a special case of the ``no warp drives'' result proven in Ref. \cite{GW00} (see Ref. \cite{sing} for a semiclassical generalization).
	\end{enumerate}
\end{enumerate}

\subsection{Definition of Maximin Surfaces}\label{def}

We now introduce the maximin construction used in the rest of this paper:

\begin{enumerate}[resume]
\item Definition: On any complete achronal slice $\Sigma$,\footnote{i.e. an AdS-Cauchy surface, using the convention in which a Cauchy surface is permitted to be null \cite{geroch70}}, where $\partial A \in \Sigma$, let the minimal area codimension 2 surface which is anchored to $\partial A$, and which is homologous to $A$, be called $\mathrm{min}(A,\Sigma)$.
	\begin{enumerate}
	\item If there are multiple minimal area surfaces on $\Sigma$, then $\mathrm{min}(A,\Sigma)$ can refer to any of them.
	\end{enumerate}

\item \label{Mdef} Definition: The maximin surface $M(A)$ is defined as the $\mathrm{min}(A,\Sigma)$ having maximal area, when varying over possible choices of complete achronal slices $\Sigma$.
	\begin{enumerate}
	\item By construction, such a maximin surface has an associated complete achronal slice $\Sigma_{M(A)}$ on which it is minimal.
	\item \label{sdef} In the case where there are multiple such surfaces, let $M(A)$ be defined as the one which is \emph{stable} in the sense that when $\Sigma$ is deformed infinitesimally to a nearby slice $\Sigma^\prime$, there still exists a surface $M^\prime(A)$ in the neighborhood of $M(A)$ with no greater area, i.e. $\mathrm{Area}[M(A)] \le \mathrm{Area}[M^\prime(A)]$.  If there are two or more stable maximin surfaces, $M(A)$ can refer to any of them.
	\end{enumerate}
\end{enumerate}
For more information about stable vs. unstable maximin surfaces, see section \ref{stable}.

\subsection{Existence of Maximin Surfaces}\label{exist}

In section \ref{hless} we will consider the case where the spacetime has no past or future global event horizons, so that signals can travel between the boundary and the interior of the bulk in a finite amount of time.  This is needed in Theorem \ref{Mexists} to prove the existence of the maximin surface $M$, since otherwise the maximization problem used to find $M$ could turn out to have solutions at infinite values of $t$ (e.g. if it touched a future singularity).  

This assumption will be weakened in section \ref{Kasner}, where horizons will be allowed so long as the only additional boundary which is introduced is a singularity which is asymptotically described by the Kasner metric (or any other homogeneous cosmology obeying the strict NCC inequality).  One may also consider spacetimes with multiple asymptotically AdS regions, and regions which involve parts of one AdS region and parts of another, as in Ref. \cite{HM13}.  It turns out that the maximin surface always avoids touching a Kasner singularity, so the maximin construction can still be performed.  Perhaps future work can generalize this to other types of singular boundaries (e.g. null singularities).

More problematic is the case where there exists an inflating region behind the horizon which has a future de Sitter boundary.\footnote{Note that the NCC requires this region to be located far enough behind the horizon that infalling observers cannot reach it \cite{FG87}}.  This will tend to make the maximin construction ill-defined, since often $\mathrm{min}(A)$ will become arbitrarily large as $\Sigma$ approaches the de Sitter boundary.  In such cases, the HRT surface need not exist either, at least in real spacetime \cite{FM14}


We start by proving the existence of a minimal area surface on a given slice:

\begin{enumerate}[resume]
\item \label{mexists} Theorem: $\mathrm{min}(A,\Sigma)$ exists.  Proof: Consider the space $\mathcal{S}$ of all codimension 2 surfaces $s$ on $\Sigma$ that are homologous to $A$.\footnote{This includes surfaces that touch the AdS boundary at points other than $\partial A$, but the maximin surface M(A) presumably cannot do so.  Since the AdS boundary is at infinite distance, and also has an infinite redshift factor for its metric, it is never efficient to gratuitously touch the AdS boundary unless $\Sigma$ and $s$ become null in the limit that they approach the boundary.  In that case, $\mathrm{min}(A,\Sigma)$ might touch the AdS boundary.  However, deforming $\Sigma$ so that it is no longer null would then increase the area of that minimal surface, so that it cannot be a maximin surface.} Each choice of $s$ must divide all other points of $\Sigma$ into two open regions, an interior region $\mathrm{Int}(s)$ and an exterior region $\mathrm{Ext}(s)$.  For ease of visualization we will refer to points in the former as black and points in the latter as white.
	\begin{enumerate}
	\item Because the spacetime is AdS-hyperbolic, all choices of $\Sigma$ have the same topology, which can be conformally compactified by including the AdS boundary.  This allows one to regard $\Sigma$ as a compact metric space, which defines a notion of distance.  \emph{Note: this is not the same as the usual geometrical distance on $\Sigma$ used to define the area.}
	\item \label{mintop} There is also a natural topology on $\mathcal{S}$ defined using the following metric: for any two surfaces $s_1$ and $s_2$, consider the upper bound of all of the following distances:
		\begin{enumerate}
		\item For each black point of $s_1$, the nearest distance to a black point of $s_2$.
		\item For each black point of $s_2$, the nearest distance to a black point of $s_1$.
		\item For each white point of $s_1$, the nearest distance to a white point of $s_2$.
		\item For each white point of $s_2$, the nearest distance to a white point of $s_1$.
		\end{enumerate}
\begin{figure}[hbt]
\centering
\includegraphics[width=.35\textwidth]{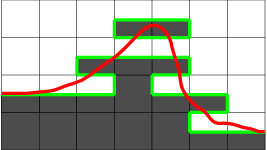}
\caption{\footnotesize Proof that $\mathcal{S}$ is compact.  After dividing $\Sigma$ into finitely many neighborhoods (shown as squares), each neighborhood may be divided into all black, all white, or half-and-half.  The boundary of the black region (shown in green) gives an approximation to the red surface.
}\label{approx}
\end{figure}
	\item $\mathcal{S}$ is compact.  To prove this it is necessary to show that any choice of $s$ can be approximated to within any precision $\epsilon$ by an element of some finite set.  Because $\Sigma$ is compact, we can divide it into a finite number $n$ of neighborhoods $\mathcal{N}$ such that any two points in the same neighborhood are no more than $\epsilon$ distance apart.  Let each neighborhood $\mathcal{N}$ be arbitrarily divided into two parts $\mathcal{N}_1$ and $\mathcal{N}_2$.  We then approximate each $s$ as follows: for each neighborhood $\mathcal{N}$, if $\mathcal{N}$ is all white or all black, we leave it the way it is, but if $\mathcal{N}$ is partly black and partly white, we paint $\mathcal{N}_1$ black and $\mathcal{N}_2$ white (see Fig. \ref{approx}).  This allows one to approximate any $s$ to accuracy $\epsilon$ with a finite number $3^n$ of points in $\mathcal{S}$.
\begin{figure}[hbt]
\centering
\includegraphics[width=.6\textwidth]{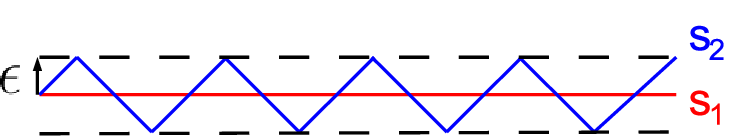}
\caption{\footnotesize The topology defined in \ref{mintop} considers two surfaces $s_1$ and $s_2$ on $\Sigma$ to be ``close'' so long as their interior (and exterior) points are within $\epsilon$ of each other, with respect to a compactified distance metric on $\Sigma$.  This allows for $s_1$ to be approximated by a surface $s_2$ which has much larger area.  By proving that $\mathrm{Area}[s_2] > \mathrm{Area}[s_1]$ in the limit of small $\epsilon$, one can show that the area is lower semicontinuous (Lemma \ref{lower}).  (A similar looking diagram can be drawn for the embedding of $\Sigma_1$ and $\Sigma_2$ in spacetime (Lemma \ref{upper}), but in that case the area is upper semicontinuous.)
}\label{jags}
\end{figure}
	\item \label{lower} $\mathrm{Area}[s]$ is lower semicontinuous.  First note that $\mathrm{Area}[s]$ is not continuous, because when two surfaces $s_1$ and $s_2$ are close in the sense of topology \ref{mintop}, then the distance between the points of $s_1$ and $s_2$ is small, but the difference between the derivatives need not be small.  Hence one can approximate a surface with low area using a limit of very jagged surfaces with large area (see Fig. \ref{jags}).  To see that the reverse is not the case, choose a normal coordinate system in which $s_1$ is located at $x = 0$ and is labelled by coordinates $i = y,\,z\ldots$ such that $g_{xi} = 0$ and $g_{xx} \ge 0$ at $s_1$, and $g_{ij}(x)$ is a continuous function.  The area of the surface $s_1$ is given by
\begin{equation}
\mathrm{Area}[s_1] = \int_{s_1} \sqrt{\mathrm{det}(g_{ij})},
\end{equation}
while the area of a nearby surface $s_2$ located at $x = f(y_i)$ is given, in the limit that $s_2$ approaches $s_1$, by
\begin{equation}\label{s_2}
\mathrm{Area}[s_2] = \int_{s_1} \sqrt{ \mathrm{det} 
\left( g_{ij} + g_{xx} \frac{\partial f}{\partial i}\frac{\partial f}{\partial j} \right) } \ge \mathrm{Area}[s_1].
\end{equation}
This inequality holds even in cases where $s_2$ backtracks so that $f$ is a multivalued ``function''.  
In cases where $s_1$ does not have a tangent defined at each point, or where it touches the AdS boundary elsewhere than $\partial A$, the area of $s_1$ may be \emph{defined} as the lower bound of all possible approximating series of surfaces.  This satisfies lower semicontinuity by definition.
	\item Hence, by the extreme value theorem, there exists an $s = \mathrm{min}(A,\Sigma)$ with minimal area.
	\item Comment: for discussion of a similar existence problem for ``outer marginally trapped'' surfaces, see Ref. \cite{plat}.
	\end{enumerate}
\end{enumerate}

\subsubsection{For Horizonless Spacetimes}\label{hless}

Next we show that on horizonless spacetimes, the maximization step can be performed:

\begin{enumerate}[resume]
\item \label{Mexists} Theorem: On a spacetime without horizons, $M(A)$ exists.  Proof: Because the spacetime is AdS-hyperbolic, its topology is given by the tensor product $X \times R_t$ where $X$ is the topology of space and $R_t$ is labelled by a time function \cite{geroch67}.  This allows any complete achronal slice $\Sigma$ to be viewed as a continuous function $t(X)$ on the conformally compact space $X$ which indicates at each spatial position the time of the slice $\Sigma$.  One can then invoke the Ascoli-Arzel\'{a} theorem \cite{AA}, which states that a subset of a space of continuous functions over a compact domain is compact with respect to the uniform topology iff it is a) equibounded, b) equicontinuous, and c) closed.  (Recall that the uniform topology comes from a metric in which the ``distance'' between two functions is the upper bound of their difference.)
	\begin{enumerate}
	\item \label{equib} The space $\mathcal{A}$ of complete achronal slices $\Sigma$ is equibounded, i.e. for each point of $X$ there is a bound $t_\mathrm{min}(X) \le t \le t_\mathrm{max}(X)$ which applies to each possible $\Sigma$.  By assumption, there are no past or future horizons in the spacetime.  This means that signals sent in at the speed of light from any point on the boundary will reach any point in the interior of $X$ in a finite amount of time.  Consequently, since $\Sigma$ is required to intersect the boundary at $\partial A$, there is a maximum and minimum value of $t$ at each point of $X$, since otherwise $\Sigma$ would not be achronal.  (In the special case where $D_A$ is the entire AdS boundary, $\partial A$ is empty, but so is $M(A)$.)
	\item $\mathcal{A}$ is equicontinuous, i.e. for each point of $X$ there is a bound on the spatial derivative of $t$ which applies to each possible $\Sigma$.  This bound comes from the lightcones at each point, since an achronal slice must always lie between the past and future lightcones.  The extreme value theorem may be used to show the existence of a nonzero minimal speed of light in the compact interval $[t_\mathrm{min}(X),\, t_\mathrm{max}(X)]$.
	\item $\mathcal{A}$ is closed, because it includes slices that are null as well as spacelike.  Hence, by the Ascoli-Arzel\'{a} theorem, the space of achronal slices is compact.
	\item \label{upper} $\mathrm{Area[min}(A,\Sigma)]$ is upper semicontinuous.  The proof is similar to \ref{lower} except that the signature of spacetime is different.  Consider a surface $s$ in $X$, let $\Sigma_1$ be an achronal slice, and let $\Sigma_2$ be a nearby slice.  Let $\Sigma_1$ be labelled by coordinates as in \ref{lower}.  Choose an additional $\tau$ coordinate which is normal to the slice $\Sigma_1$, so that $g_{\tau_x} = g_{\tau i} = 0$ and $g_{\tau\tau} \le 0$.   The area of $s$ on the slice $\Sigma_1$ is given as before by
\begin{equation}
\mathrm{Area}[s, \Sigma_1] = \int_{s} \sqrt{\mathrm{det}(g_{ij})},
\end{equation}
while the area of $s$ on the nearby slice $\Sigma_2$ located at $\tau = f(y_i)$ is given, in the limit that $\Sigma_2$ approaches $\Sigma_1$, by
\begin{equation}\label{sigma_2}
\mathrm{Area}[s, \Sigma_2] = \int_{s} \sqrt{ \mathrm{det} 
\left( g_{ij} + g_{\tau\tau} \frac{\partial f}{\partial i}\frac{\partial f}{\partial j} \right) } \le \mathrm{Area}[s, \Sigma_1].
\end{equation}
Now, if the area of each slice $s$ is upper semicontinuous, then the area of the minimum $\mathrm{min}(A,\Sigma)$ must also be.  

In cases where $\Sigma_1$ does not have a tangent defined at each point, the area of $s$ at $\Sigma_1$ may be \emph{defined} as the upper bound of $\mathrm{min}(A,\Sigma)$ on all possible approximating series of slices.  This satisfies upper semicontinuity by definition.
	\item Hence, by the extreme value theorem, there exists a $\Sigma$ such that $\mathrm{min}(A,\Sigma) = M(A)$ has maximal area.
	\end{enumerate}
\end{enumerate}

\subsubsection{For Black Holes with Kasner Singularities}\label{Kasner}

The restriction to horizonless spacetimes is a serious constraint, because often we are interested in HRT surfaces which pass through black hole regions, e.g. \cite{HM13}.

So suppose now that the spacetime does have black hole horizons, but that any past or future boundaries which $\Sigma$ might touch are singularities governed by Kasner-like behavior.  That is, we assume that the singularities are spacelike and that the metric near the singularity asymptotically approaches the following Kasner-like form:
\begin{equation}
ds^2 = -dt^2 + \sum_{i=0}^{D-1} t^{2p_i} dx_i^2
\end{equation}
where $\sum_i p_i = \sum_i p_i^2 = 1$, and $t = 0$ at the singularity.  (If one $p_i = 1$ and the rest vanish, we have the trivial Kasner solution, which is just a patch of flat Minkowski space in Milne-like coordinates.) This metric is valid e.g. in the deep interior of a Schwarzschild black hole, and also during each Kasner phase of a chaotic BKL singularity \cite{BKL70}.  In this situation, we can also prove the existence of maximin surfaces:

\begin{enumerate}[resume]
\item\label{Kas} Theorem: $M(A)$ also exists on a spacetime with horizons, so long as the only singularities (that a slice $\Sigma \ni \partial A$ could touch) are spacelike, and the metric near them takes on a nontrivial Kasner form. 

	\begin{enumerate}
	\item The proof of \ref{Mexists} goes through without change, except for step \ref{equib} in which the absence of horizons was used.  The only thing that could go wrong is if, when maximizing the area of $\mathrm{min}(A,\Sigma)$, the maximum surface ends up touching the singularity.  In order to rule this out, we must show that as $\Sigma$ approaches the singularity, $\mathrm{Area}[\mathrm{min}(A,\Sigma)]$ decreases rather than increases.  Let us assume for contradiction that the maximal choice of $\Sigma$ approaches the singularity in some neighborhood $\mathcal{N}$, in which the Kasner solution is valid, and let $t_\mathrm{min}[\Sigma]$ be the time closest to the singularity on $\Sigma$.  We shall show that it is always better to adjust $\Sigma$ to a $\Sigma^\prime$ such that $t_\mathrm{min}[\Sigma^\prime] > t_\mathrm{min}[\Sigma]$

	\item Let us begin by calculating the area of a codimension 2 surface $s$ situated at a fixed
location with respect to the $x$ coordinates, at a particular choice of time $t$.  Let us order the $D-1$ $p_i$ values, so that $p_0 < p_1 \ldots < p_{D-1}$.  In order to show that this area is always decreasing as we move towards the singularity, consider the worst case scenario in which the codimension 2 surface is located along the first $D-2$ directions.  Its area is therefore given by:
\begin{equation}
\mathrm{Area}[s] = \prod_{i = 0}^{D-2} t^{p_i} = t^{1 - p_{D-1}}
\end{equation}
using the linear sum rule.  However, the quadratic sum rule tells us that $p_{D-1} < 1$ for nontrivial Kasner.  Hence $\lim_{t \to 0} \mathrm{Area}[s] = 0$.  It follows that so long as $\Sigma$ is located at a fixed Kasner time $t$ within $\mathcal{N}$, the maximization procedure tells us to take $t$ large in order to avoid the singularity.
	\item In general $\Sigma$ is given by a nonconstant time function $t(x)$.  However, when $\Sigma$ is slanted at a boost, this only decreases the amount of area for any given $s$.  Hence, if we define a new $\Sigma^\prime$ within $\mathcal{N}$ by
\begin{equation}
t(x)_{\Sigma^\prime} = \mathrm{max}[t(x)_\Sigma,\,T], \qquad T > t_\mathrm{min}[\Sigma],
\end{equation}
the area of all surfaces $s$ are increased by this deformation.
	\item If \emph{each} surface $s$ has increasing area as you move away from the singularity, the same holds for $\mathrm{Area}[\mathrm{min}(A,\Sigma)]$.
	\item Since this argument could be repeated for any other slice $\Sigma$ passing through $\mathcal{N}$, it follows that $M(A)$ cannot be located in any neighborhood $\mathcal{N}$ where a nontrivial Kasner solution is valid.  Hence there is a barrier at a finite time from the singularity, beyond which no maximin surface can pass \cite{barriers}.
	\item Corollary: The same result applies to other anisotropic singularities, so long as any $D-1$ directions are contracting on average.\footnote{At least for homogeneous power-law cosmologies of the type being considered, this automatically follows so long as the null curvature $R_{kk}$ is positive, as can be seen through use of the Raychaudhuri equation (\ref{Ray}).} This includes isotropic FRW-type singularities where all spatial directions are contracting (due to the presence of matter).
	\end{enumerate}
\end{enumerate}
From the considerations above, it appears that there may be trouble with the maximin conjecture in cases where there are inflating de Sitter boundaries behind a horizon.  For in this case, the area of a codimension 2 surface will tend to increase as one approaches the de Sitter boundary.  In certain cases, this may make it so that maximin/extremal surfaces fail to exist \cite{FM14}.

\subsection{Stability Issues}\label{stable}

According to definition \ref{sdef}, the maximin surface $M(A)$ is required to be stable, meaning that for any way of slightly deforming the slice $\Sigma_{M(A)}$, there still exists a nearby minimal area surface.  Unstable maximin surfaces can arise when $\Sigma_{M(A)}$ happens to pass through another surface with the same area, but they are not particularly interesting (for example, they need not be extremal).  Hence, after this section we will use the term ``maximin'' to refer to the stable case only.

\begin{figure}[hbt]
\centering
\includegraphics[width=.7\textwidth]{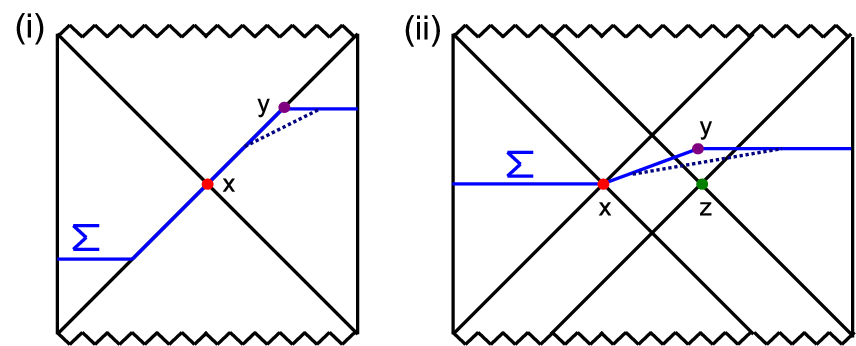}
\caption{\footnotesize Examples of unstable maximin surfaces: (i) An eternal static AdS-Schwarzschild black hole.  The bifurcation surface $x$ is a maximin surface, and is also extremal.  However, technically, \emph{any} other slice $y$ of the past or future horizon is also a maximin surface, because there exist achronal slices $\Sigma$ which follow the horizon, and $\mathrm{Area}(x) = \mathrm{Area}(y)$.  These other slices $y$ are unstable because if $\Sigma$ is slightly deformed (as shown by the dotted line), there is no longer a minimal area surface near $y$.  These unstable maximin surfaces disappear when the horizons are not exactly static.  (ii) A wormhole with two extremal surfaces $x$, $z$ in its throat, such that $\mathrm{Area}(z) > \mathrm{Area}(x) = \mathrm{Area}(y)$.  Once again, $x$ is the stable maximin surface, but $y$ is an unstable surface since $\Sigma$ can be chosen to pass through $y$.  In neither case (i) nor (ii) is $y$ extremal.}\label{unstable}
\end{figure}

Some examples of unstable maximin surfaces are shown in Fig. \ref{unstable}.

\begin{enumerate}[resume]
\item \label{sta} Theorem: If the maximin surface $M(A)$ is the unique minimum area surface on $\Sigma_{M(A)}$, then it is stable. If there are two or more maximin surfaces on $\Sigma_{M(A)}$, then at least one of them is stable, at least if $\Sigma_{M(A)}$ is smooth and spacelike.
	\begin{enumerate}
	\item In the case where there is only one maximin surface $M(A)$, it is necessarily stable.  For suppose it were unstable under a variation to some slice $\Sigma^\prime$, then $\mathrm{min}(A,\Sigma^\prime)$ would have greater area than $M(A)$, contradicting the maximality of $M(A)$.
	\item At least when $\Sigma_{M(A)}$ is spacelike and smooth, if there are multiple maximin surfaces $M_1(A),\,M_2(A)\ldots$, then at least one of them is stable.  We start by showing that $M_1(A),\,M_2(A)\ldots$ are all disjoint.  For if not, there would be two minimal surfaces lying on the same smooth $\Sigma$ which either touch or cross each other.  
	\item Suppose two minimal surfaces $M_1$ and $M_2$ cross each other, dividing each other into segments $M_1 = ab$ and $M_2 = cd$.  Then it is possible to reconnect these segments into two other surfaces $M_3 = ad$ and $M_4 = cb$ (see Fig. \ref{squiggle}).  If $\mathrm{Area}[b] < \mathrm{Area}[d]$, then $\mathrm{Area}[bc] < \mathrm{Area}[cd]$ which contradicts the minimalness of $M_2$.  If $\mathrm{Area}[b] > \mathrm{Area}[d]$, then $\mathrm{Area}[ad] < \mathrm{Area}[ab]$ which contradicts the minimalness of $M_1$.  The remaining possibility is $\mathrm{Area}[b] = \mathrm{Area}[d]$, but in this case $ad$ and $cb$ would have to be minimal surfaces, which is impossible since they have sharp corners (whose area could be decreased by rounding them off).
\begin{figure}[hbt]
\centering
\includegraphics[width=.5\textwidth]{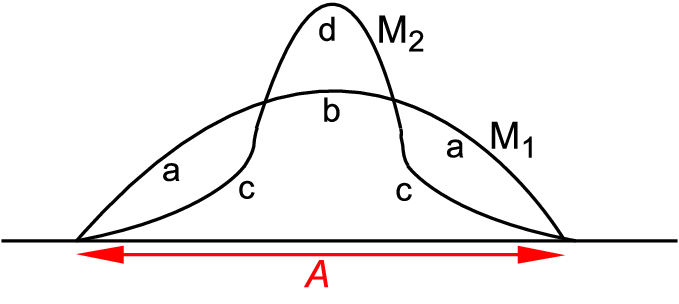}
\caption{\footnotesize The maximin surfaces $M_1$ and $M_2$ are each minimal surfaces on $\Sigma$ anchored to $\partial A$.  They cannot cross each other, or else one of $ad$ or $bc$ would have area no greater than $M_1$ or $M_2$ respectively.}\label{squiggle}
\end{figure}
	\item Suppose instead that $M_1$ and $M_2$ touch each other without crossing.  Otherwise they would have to curve away from each other in the neighborhood of that point.  By Corollary \ref{Kineq}, at some point $y$ near $x$, the extrinsic curvatures would satisfy 
$\mathrm{tr}(K)[M_2] \ne \mathrm{tr}(K)[M_1]$.  But then the two surfaces cannot both satisfy $\mathrm{tr}(K) = 0$ as needed to be a minimum area surface.  Hence $M_1(A),\,M_2(A)\ldots$ are all disjoint.
	\item Suppose now that $M_1(A),\,M_2(A)\ldots$ are all unstable.  This means that for each $M_n$, it would be possible to vary $\Sigma$ to $\Sigma^\prime$ in a neighborhood of $M_n(A)$, in such a way that the area of each nearby $M_n^\prime(A)$ increases.  But because each of the $M_n(A)$'s are disjoint, it is possible to perform all of these variations simultaneously.  This gives us a slice $\Sigma^\prime$ for which the area of every surface anchored to $\partial A$ is greater than that of $M(A)$.  But that contradicts the maximality property of the maximin surfaces.  It follows that at least one of $M_1(A),\,M_2(A)\ldots$ are stable under perturbations.
	\end{enumerate}
\end{enumerate}
Ideally, it would be desirable if this theorem could be proven when $\Sigma$ is an arbitrary achronal surface, not just smooth and spacelike.  If so, it could be combined with Theorem \ref{Mexists} or \ref{Kas} to prove the existence of at least one stable maximin surface.  

It would be good to fill this lacuna, in order to obtain complete proofs of e.g. strong subadditivity for the HRT surfaces.  However, in most contexts where one would apply HRT, it is a reasonable assumption that stable maximin surfaces exist.  In light of this, we will proceed under the assumption that stable maximin surfaces can always be found.  This will enable us to prove that the stable maximin surface is in fact none other than the HRT surface.

\subsection{Equivalence of Maximin and HRT Surfaces}\label{equiv}

In this section we will show that the maximin surface is identical to the HRT surface.  The basic intuition is fairly simple: the maximin surface is minimal in one of its two normal directions, and maximal in another.  Since a codimension 2 surface can only be varied in two directions, this means it has to be an extremal surface.

Since the devil is in the details, we need to work our way up to this result more gradually.  First we need a definition of tangent vectors which works even in cases where $\Sigma$ is jagged.  Then, we need to show that $M(A)$ can in fact be freely varied in all of its normal directions.  Since $M(A)$ is constrained to lie on an achronal slice $\Sigma$, this requires showing that no two points on $M(A)$ are null separated (this requires the NCC).  Only then will we be able to prove the equivalence to the HRT surface (using the NCC once again).

\begin{enumerate}[resume]
\item Definition: A vector $\mathbf{v}$ is a tangent vector of $\Sigma$ at point $x$ iff some points on $\Sigma$ near $x$ lie at an arbitrarily small angle to $\mathbf{v}$.  More precisely, let $\mathcal{V}$ be the space of spacetime vectors at $x$ modulo multiplication by positive reals.  $\mathcal{V}$ can be given the topology of a $(D-1)$-sphere.  Let $\mathcal{N}$ be a neighborhood of $x$, and let $y$ be some point on $\Sigma \cap \mathcal{N}$ connected to $x$ by a geodesic ray $g$ lying in $\mathcal{N}$.  Let $\mathbf{w}$ be a vector at $x$ pointing along $g$ towards $y$.  If, even when $\mathcal{N}$ is taken to be small, there always exist points $y$ for which $\mathbf{v}$ and $\mathbf{w}$ are arbitrarily close in $\mathcal{V}$, then $\mathbf{v}$ is a tangent vector.
	\begin{enumerate}
	\item Because $\Sigma$ may have a discontinuous first derivative, one cannot assume that if $\mathbf{v}$ is a tangent vector, $-\mathbf{v}$ will also be a tangent vector.
	\item At a point where a two dimensional surface $P$ with Lorentzian signature intersects $\Sigma$, there always exist at least two distinct vectors in $\mathcal{V}$ which are tangent to $\Sigma$, one on either side.  However, if $\Sigma$ is sufficiently wiggly, there may be more on either side (e.g. if $\Sigma$ were like the function $f(x) = x \sin x$).
	\item The tangent vectors of $\Sigma$ indicate the directions in which one can vary the location infinitesimally, while remaining on $\Sigma$.
	\end{enumerate}

\item \label{notnull} Theorem: $M(A)$ is everywhere spacelike separated to itself, i.e. no two points are null separated.  If $\Sigma_{M(A)}$ is a spacelike slice, this is obvious.  But what if $M(A)$ is minimal on a $\Sigma$ which is partly null?
	\begin{enumerate}
	\item \label{segnull} No segment of $M(A)$ is null.  For suppose there were one or more null segments $n$ on $M(A)$.  Define $k^a$ as a future-null tangent vector on $n$.  Then let $\Sigma = t(X)$ be slightly deformed into an everywhere spacelike slice $\Sigma^\prime = t(X) + \epsilon f(X)$, such that $\epsilon$ is an infinitesimal parameter, $f(X)$ has support only in a small neighborhood of some $n$, and $k^a \nabla_a f(X) < 0$ (as needed to make $\Sigma^\prime$ become spacelike).  Then each line segment $n$ on $\Sigma$ is deformed into a segment $n^\prime$ on $\Sigma^\prime$ having a positive length of order $\sqrt{\epsilon}$.  All other changes in lengths are of order $\epsilon$, which are negligible in comparison.  Consequently $\mathrm{Area}[\mathrm{min}(A,\Sigma^\prime)] > \mathrm{Area}[\mathrm{min}(A,\Sigma]$,
which contradicts the assumption that $M(A)$ is maximal when varying $\Sigma$.
	\item Neither can two points on $M(A)$ be connected by a null segment $n$ which does not lie on $M(A)$ (see Fig. \ref{nulls}).   Let $x$ be the point to the past, and $y$ the point to the future.  Let us shoot out a null congruence $N_x$ from $M(A)$ in a neighborhood of the point $x$.   Choose an affine parameter on $n$ whose unit vector $k^a$ points towards the future.  Using the Raychaudhuri equation, the NCC, and the generic condition on $N_x$, it follows that either $\theta(N_x) > 0$ at $x$, or else $\theta(N_x) < 0$ at $y$.  $N_x$ does not necessarily contain $M(A)$ near the point $y$, but if it does not, $M(A)$ must bend outwards away from $N_x$.\footnote{In this theorem, ``outwards'' means away from $n$, towards the periphery of Fig. \ref{nulls}.}  Therefore the null congruence $N_y$ shot out from $M(A)$ near $y$ will have $\theta(N_y) \le \theta(N_x)$.  So either $\theta(N_x) > 0$ or $\theta(N_y) < 0$.  By symmetry, we select the former case, in order to derive a contradiction.
\begin{figure}[hbt]
\centering
\includegraphics[width=.5\textwidth]{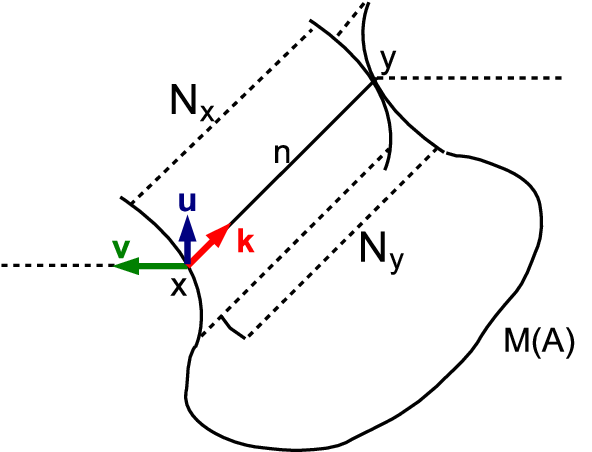}
\caption{\footnotesize Two points $x$ and $y$ on $M(A)$ are separated by a null geodesic $n$ in $\Sigma$.  $N_x$ is a null congruence shot out from the points of $M(A)$ near $x$.  Behind it is the null congruence $N_y$ shot out from $M(A)$ near $y$.  Comparison of these null congruences allows one to place constraints on the sign of the expansion $\theta$ at $x$ or $y$.  $k^a$ and $v^a$ are tangent vectors of $\Sigma$ at $x$, along which $M(A)$ is minimal.  But $u^a$ is a vector pointing to the future of $\Sigma$, along which $M(A)$ is maximal---a contradiction.}\label{nulls}
\end{figure}
	\item  Consider the trace of the extrinsic curvature at $x$:
\begin{equation}\label{Ki}
\mathrm{tr}(K)[M(A)] = (\mathrm{Area})^{-1} \nabla \mathrm{Area} \equiv K_i,
\end{equation}
regarded as a tangent covector pointing in the 2 dimensional spacetime plane normal to $M(A)$.  By the preceding argument, the trace of the null extrinsic curvature $\theta(N_x) = K_i k^i > 0$ along $n$ is positive.  There must be at least one other tangent vector $v^i$ of $\Sigma$, besides $k^i$.  Since $M(A)$ is minimal on $\Sigma$, it follows that $K_i v^i \ge 0$.  Hence $v^i$ and $k^i$ are not diametrically opposed.  $v^i$ must therefore point in an outward-spacelike direction, since $k^i$ points in an inward-null direction.
	\item Let $\Sigma$ now be pushed infinitesimally to the future, into a new slice $\Sigma^\prime$.  By the maximality property, $\mathrm{Area}[\mathrm{min}(A,\Sigma^\prime)] \le \mathrm{Area}[M(A)]$.  This implies that there exists a vector $u^i$ pointing into the future of $\Sigma$ for which $K_i u^i \le 0$.  But $u^i$ is a positive linear combination of $v^i$ and $k^i$, so $K_i u^i > 0$ which is a contradiction.  
	\item Corollary: In the case where there are multiple maximin surfaces, no two points on any $M(A)$'s can be null separated from each other.
	\end{enumerate}

\item \label{Mism} Theorem: A maximin surface is an HRT surface (and conversely): $M(A) = m(A)$.
	\begin{enumerate}
	\item \label{ext1} $M(A)$ is extremal.  Proof: First we consider the case where $\Sigma_{M(A)}$ has a continuous first derivative.  At each point of $M(A)$, one can vary the location of $M(A)$ in the two spacetime dimensions normal to $M(A)$.  Since $M(A) = \mathrm{min}(A,\Sigma)$ for $\Sigma = \Sigma_{M(A)}$, it is minimal (and hence extremal) with respect to variations along $\Sigma$.  By varying $\Sigma$ with time as a function of position, the maximality property guarantees that it is maximal (and hence extremal) in at least one additional direction.  (Theorem \ref{notnull} implies that there is no obstruction in varying $\Sigma$ to the past and future.)  By linearity of first order variations, this is sufficient to prove extremality in any other direction as well.
	\item \label{ext2} In the case where the first derivative of $\Sigma$ jumps discontinuously, $M(A)$ must still be extremal.  Let $\mathbf{v}$ and $\mathbf{w}$ be the tangent vectors of $\Sigma$ in the plane normal to $M(A)$, along which $M(A)$ is minimal.  Let $\mathbf{p}$ and $\mathbf{q}$ be the vectors along which $M(A)$ is maximal, one pointing to the future of $\Sigma$, the other to the past.  If $\mathbf{v}$ and $\mathbf{w}$ are not diametrically opposed, then either $\mathbf{p}$ or $\mathbf{q}$ is a positive linear combination of $\mathbf{v}$ and $\mathbf{w}$.  This is only possible if $M(A)$ is extremal along all three of $\mathbf{p}$ (or $\mathbf{q}$), $\mathbf{v}$ and $\mathbf{w}$.  Since this includes two linearly independent directions, $M(A)$ is extremal.  Similarly if $\mathbf{p}$ and $\mathbf{q}$ are not diametrically opposed.  The remaining case, where each pair of vectors is diametrically opposed, is extremal by Lemma \ref{ext1}.
	\item \label{RT1} $M(A)$ has less area than any extremal surface $x(A)$ which does not lie on $\Sigma$.  Proof: Consider $x(A)$'s representative $\tilde{x}(A,\Sigma)$ (which exists and is homologous to $A$ by Corollary \ref{repex}).  It follows that $\mathrm{Area}[M(A)] \le \mathrm{Area}[\tilde{x}(A)] < \mathrm{Area}[x(A)]$ where the first inequality uses minimality of $M(A)$ on $\Sigma$ and the second uses Theorem \ref{trap}.
	\item \label{RT2} By construction, $M(A)$ has no more area than any other extremal surface on $\Sigma$.  Hence $M(A)$ is an HRT surface $m(A)$.  Conversely, if there is another HRT surface on $\Sigma$, it has minimal area on $\Sigma$, and hence is also maximin.  So the two definitions are equivalent.
	\item Corollary: $M(A)$ is not minimal on every slice $\Sigma^\prime$ that passes through $M(A)$.  For let $\Sigma^\prime$ be a null congruence $N(A)$ shot out from $M(A)$.  Because $M(A)$ is extremal, $\theta = 0$, but by Theorem \ref{trap}, 
$d \theta / d \lambda < 0$.  Hence $M(A)$ is locally a maximum of area rather than a minimum.  By continuity, the same applies if $\Sigma^\prime$ is a highly boosted spacelike slice.
%
	\end{enumerate}
\end{enumerate}

\subsection{Properties of Maximin/HRT Surfaces}\label{props}

Now we use the maximin construction to prove some nice properties of the maximin/HRT surfaces: they have less area than the causal surface (\ref{LESS}), move outward monotonically as the boundary region grows (\ref{GROW}), obey strong subadditivity (\ref{STR}), and also monogamy of mutual information (\ref{MONO}).

\subsubsection{Less Area than the Causal Surface}\label{LESS}

First we show that the causal surface has more area than the extremal surface, as postulated by Hubeny and Rangamani \cite{HR12}.  This suggests that $\mathrm{Area}[w(A)]$ might be a coarse-grained measure of the entropy in $A$, as proposed in Refs. \cite{HR12,coarse}.  This proof uses the NCC.
\begin{figure}[hbt]
\centering
\includegraphics[width=.4\textwidth]{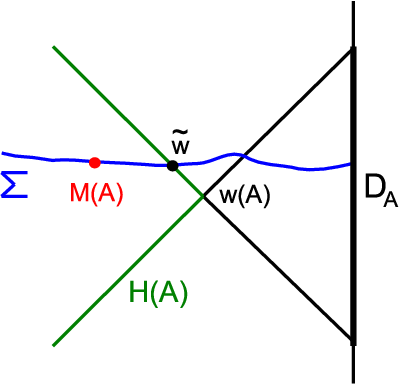}
\caption{\footnotesize $H(A)$ is the intersection of the past and future horizon continued beyond the causal surface $w(A)$.  Because the slice on which $M(A)$ is minimal intersects $H(A)$ at $\tilde{w}$, by the Second Law, $M(A)$ has less area than $w(A)$.}\label{lesser}
\end{figure}

\begin{enumerate}[resume]
\item Theorem: The area of $w(A)$ is greater than the area of $m(A)$.  Proof (see Fig. \ref{lesser}): Let $m(A) = M(A)$ be minimal on the slice $\Sigma$.  Let $H(A)$ be defined as the surface obtained by shooting out null rays from $w(A)$ in the direction away from the boundary.  $H(A)$ consists of two branches: i) the component of the future horizon $\partial I^-(D_A)$ which is to the past of $w(A)$, and ii) the component of the past horizon $\partial I^-(D_A)$ which is to the future of $w(A)$.  By the Second Law of horizons, the area of $H(A)$ decreases when moving away from $w(A)$.  Let the representative $\tilde{w}(A, \Sigma)$ be defined as $H(A) \cap \Sigma$.  Then $\mathrm{Area}[w(A)] > \mathrm{Area}[\tilde{w}(A, \Sigma)] > \mathrm{Area}[m(A)]$.  The inequalities are strict because of the generic condition.
	\begin{enumerate}
	\item Comment: It might be that $\mathrm{Area}[w(A)] - \mathrm{Area}[m(A)] = +\infty$ due to divergences near the boundary.  In that case, the area inequality should apply only to the \emph{leading order} divergence of the area difference, not necessarily to the finite quantities extracted from some renormalization scheme.

Freivogel and Mosk \cite{FM13} have shown that the subleading divergences of the causal surface $\mathrm{Area}[w(A)]$ are indeed different from $\mathrm{Area}[m(A)]$.  (In fact, the causal surface divergences can even be nonlocal on the boundary $\partial A$!)  This suggests that an infinite area difference is probably the generic situation.

	\item Corollary: The theorem applies not only to $w(A)$, but more generally to any surface $s(A)$, so long as its extrinsic curvature vector $K_i$ (defined in Eq. (\ref{Ki})) is spacelike or null, and pointing towards $D_A$ everywhere.  (One might worry that the null surface shot out from $s(A)$ might intersect the boundary at $D_{\bar{A}}$ and disappear.  However, the argument of Theorem \ref{wm} can be used to show that the surface $s(A)$ does not lie within the causal wedge
$I^-(D_{\bar{A}}) \cap I^+(D_{\bar{A}})$ of the complementary region.)
	\end{enumerate}
\end{enumerate}

\subsubsection{Moves outwards as the Boundary Region Grows}\label{GROW}

It makes intuitive sense that the larger the CFT region is, the larger the corresponding bulk region should be.  But actually proving it is tricky, and our proof here requires the full power of the maximin construction.  In the proof below, we need to use the maximin construction to simultaneously construct \emph{multiple} maximin surfaces using the \emph{same} slice $\Sigma$.  Fortunately, this appears to be possible, and gives the same results as constructing the two maximin surfaces separately.

\begin{enumerate}[resume]
\item \label{outward} Theorem: If $A \supset B$, then $r(A) \supset r(B)$, with $m(A)$ spacelike to $m(B)$.  In other words, as the region on the boundary gets bigger, $m$ has to move outwards in a spacelike direction (although some connected components of $m(A)$ and $m(B)$ may coincide exactly).  Furthermore, $m(A)$ and $m(B)$ are minimal on the same slice $\Sigma$.  Proof: We will construct a pair of maximin surfaces $M_1$ and $M_2$, by maximizing and minimizing the sum $Z = c_1 \mathrm{Area}[M_1] + c_2 \mathrm{Area}[M_2]$ on the same slice $\Sigma$, such that $M_1$ is anchored to $\partial A$ and $M_2$ is anchored to $\partial B$.  Here $c_1$ and $c_2$ are arbitrary positive coefficients (one of which could be infinitesimal compared to the other).  Despite the fact that these two surfaces are constrained to lie on the same achronal slice, nevertheless it can be shown that they are the HRT surfaces $m(A)$ and $m(B)$ respectively.
	\begin{enumerate}
	\item $M_1$ and $M_2$ exist by the same argument as Theorems \ref{Mexists} or \ref{Kas}.  Taken separately, each one is a minimal area surface on the same slice $\Sigma$.  Their points cannot be null separated, by the argument of Theorem \ref{notnull}.
\begin{figure}[hbt]
\centering
\includegraphics[width=.5\textwidth]{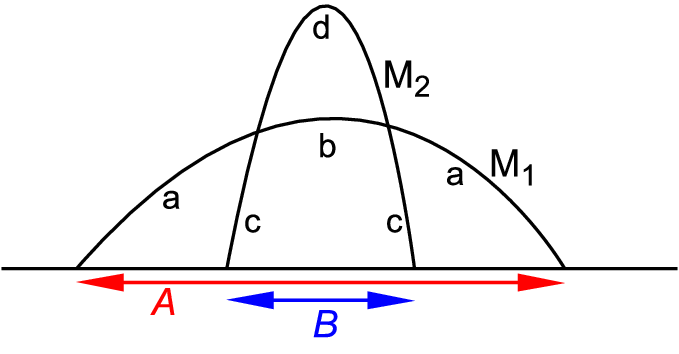}
\caption{\footnotesize The maximin surfaces $M_1$ and $M_2$ are minimal surfaces on $\Sigma$, the former anchored to $\partial A$ and the latter to $\partial B$.  They cannot cross each other, or else one of $ad$ or $bc$ would have area no greater than $M_1$ or $M_2$ respectively.  (This situation is similar to Fig. \ref{squiggle}, but here the surfaces are not anchored at the same points.)
}\label{cross}
\end{figure}
	\item \label{crossing} $M_1$ must lie spatially outside or on $M_2$.  Otherwise, there would be two minimal area surfaces on $\Sigma$ which cross each other.  If they cross each other, you can write $M_1 = ab$ and $M_2 = cd$ where $a$ is outside of $c$ but $d$ is outside of $b$ (See Fig \ref{cross}).  But then there also exist continuous codimension 2 surfaces $ad$ and $cb$.  If $\mathrm{Area}[b] < \mathrm{Area}[d]$, then $\mathrm{Area}[bc] < \mathrm{Area}[cd]$ which contradicts the minimalness of $M_2$.  If $\mathrm{Area}[b] > \mathrm{Area}[d]$, then $\mathrm{Area}[ad] < \mathrm{Area}[ab]$ which contradicts the minimalness of $M_1$.  The remaining possibility is $\mathrm{Area}[b] = \mathrm{Area}[d]$, but in this case $ad$ and $cb$ would have to be minimal surfaces, which is impossible since they have sharp corners (whose area could be decreased by rounding them off).
	\item $M_1$ and $M_2$ cannot exactly coincide, because they are anchored to different points on the boundary.  However, individual connected components of $M_1$ and $M_2$ may coincide.  In fact, if $M_1$ and $M_2$ coincide in the neighborhood of any point $x$, then their connected components that include $x$ must coincide everywhere.  For as minimal surfaces their position is the solution to the elliptical equation $\mathrm{tr}(K) = 0$.  By Corollary \ref{elliptical}, they must therefore coincide on those connected components. 
	\item Except on connected components which exactly coincide, $M_1$ cannot touch $M_2$ at any point $x$.  Otherwise they would have to curve away from each other in the neighborhood of that point.  By Corollary \ref{Kineq}, at some point $y$ near $x$, the extrinsic curvatures would satisfy 
$\mathrm{tr}(K)[M_2] > \mathrm{tr}(K)[M_1]$ (with the sign of $K$ defined so that surfaces with positive extrinsic curvature deviate towards the AdS boundary).  But then the two surfaces cannot both satisfy $\mathrm{tr}(K) = 0$ as needed to be a minimum area surface.
	\item Hence we can divide the connected components of $M_1$ and $M_2$ into two classes, those which entirely coincide and those which do not even touch.  Restricting consideration to the coinciding components, $\mathrm{Area}[M_1] = \mathrm{Area}[M_2] \propto Z$.  This must continue to hold even when varying $\Sigma$ in a local neighborhood.  Hence for these components maximizing $Z$ is equivalent to maximizing both $\mathrm{Area}[M_1]$ and $\mathrm{Area}[M_2]$.
	\item In the case of the components of $M_1$ and $M_2$ that do not touch, it is possible to freely vary $\Sigma$ in a neighborhood of $M_1$ without affecting the area of $M_2$.  We can now apply the same argument as Theorem \ref{Mism} to show that $M_1$ and $M_2$ are the HRT surfaces $m(A)$ and $m(B)$.  By Lemmas \ref{ext1} and \ref{ext2}, $M_1$ and $M_2$ are each extremal surfaces.

Furthermore, if $(M_1, M_2)$ were not the HRT surfaces, then some other pair $(x_1, x_2)$ of extremal surfaces with lesser weighted area $Z$ would be HRT.  That would require either $\mathrm{Area}[M_1] > \mathrm{Area}[x_1]$, or else $\mathrm{Area}[M_2] > \mathrm{Area}[x_2]$.  In the former case, Lemma \ref{RT1} states that the representative $\tilde{x}_1$ on $\Sigma$ would have even less area than $M_1$, contradicting the minimality of $M_1$.  The same applies in the latter case, so $M_1$ and $M_2$ are HRT.
	\item Corollary: If there are multiple $m(A)$'s or $m(B)$'s, all of them lie on the same slice $\Sigma$ and each of the $m(B)$'s is outside each of the $m(A)$'s.
	\item \label{multiR} Corollary: For any set of disjoint spacelike-separated regions $R_n$, $n \in 1...N$, all the $m(R_i)$'s are minimal on the same slice $\Sigma$.  One can construct surfaces $M_n$ by maximizing and minimizing a quantity $Z = \sum c_n \mathrm{Area}[M_n]$.  Because of the disjointness condition, Lemma \ref{crossing} applies to any pair of $M_n$'s.  The proof that the $M_n$'s are HRT proceeds in the same way as above.
	\end{enumerate}
\end{enumerate}
In addition to being interesting in its own right, this theorem is a critically important step in the proof of strong subadditivity in the next section.

\subsubsection{Strong Subadditivity}\label{STR}

We now come to the proof of strong subadditivity.  The basic idea here to use Theorem \ref{outward} to show that the intersection and union entropies are minimal on a common slice $\Sigma$.  The other two extremal surfaces need not lie on $\Sigma$, but they have \emph{representatives} on $\Sigma$, and this is good enough, since the NCC\footnote{A. Prudenziati has pointed out to me that this use of the NCC is inessential, since one may use the minimum area surfaces $\text{min}(AB, \Sigma)$ and $\text{min}(BC,\Sigma)$ in place of $\tilde{m}(AB,\Sigma)$ and $\tilde{m}(BC,\Sigma)$ below, which have less area than $m(AB)$ and $m(BC)$ by the maximin property.  However, the NCC still enters into the proof of SSA via Theorems 6 and 14(b), which are needed to prove Theorem 15, and therefore also Theorem 17.  A similar statement applies to Theorem \ref{mono} below.} says that their representatives have lesser area.

\begin{enumerate}[resume]
\item \label{SSA} Theorem: Let $A$, $B$ and $C$ be disjoint boundary regions (which may share a boundary).  Then the Strong Subadditivity property holds:
\begin{equation}
\mathrm{Area}[m(AB)] + \mathrm{Area}[m(BC)] \ge \mathrm{Area}[m(ABC)] + \mathrm{Area}[m(B)].
\end{equation}
	\begin{enumerate}
	\item By Theorem \ref{outward}, there exists a spacelike slice $\Sigma$ on which both $m(ABC)$ and $m(B)$ lie as minimal surfaces, with $m(ABC)$ everywhere outside of $m(B)$.  
	\item By Corollary \ref{repex}, $m(AB)$ and $m(BC)$ have homologous representatives $\tilde{m}(AB,\Sigma)$ and $\tilde{m}(BC,\Sigma)$. By Theorem \ref{trap}, they have less area than $m(AB)$ and $m(BC)$ respectively (unless they coincide with them).
	\item $\mathrm{Area}[\tilde{m}(AB,\Sigma)] + \mathrm{Area}[\tilde{m}(BC,\Sigma)] \ge \mathrm{Area}[m(ABC)] + \mathrm{Area}[m(B)]$, by the static argument for strong subadditivity of minimal area surfaces shown in Fig. \ref{SSAfig} or Ref. \cite{HT07}.
	\item Corollary: Strong Subadditivity can be saturated only if $m(AB)$ and $m(BC)$ are minimal on the same slice $\Sigma$, but do not cross one another.  This requires that $B = B_1 \cup B_2$ such that $m(ABC) = m(AB_1) \cup m(B_2C)$.  
	\end{enumerate}
\end{enumerate}

\subsubsection{Monogamy of Mutual Information}\label{MONO}

The proof of the monogamy of mutual information is almost exactly the same:

\begin{enumerate}[resume]
\item \label{mono} Theorem: Let $A$, $B$ and $C$ be disjoint boundary regions (which may share a boundary).  Then the monogamy of mutual information holds:
\begin{align}
\mathrm{Area}[m(AB)] + \mathrm{Area}[m(BC)] + \mathrm{Area}[m(AC)] \ge \phantom{Mn} \notag \\ 
\mathrm{Area}[m(A)] + \mathrm{Area}[m(B)] + \mathrm{Area}[m(C)]  + \mathrm{Area}[m(ABC)].
\end{align}
	\begin{enumerate}
	\item By Corollary \ref{multiR}, there exists a spacelike slice $\Sigma$ on which $m(A)$, $m(B)$, $m(C)$ and 
$m(ABC) = m(\overline{ABC})$ are all minimal.
	\item By Corollary \ref{repex} and Theorem \ref{trap}, $m(AB)$, $m(BC)$, and $m(AC)$ all have homologous representatives on $\Sigma$ which have less area (unless they coincide with them).
	\item One can then apply the static proof of Monogamy found in Ref. \cite{HHM11}.  This involves chopping each of $m(AB)$, $m(BC)$, and $m(AC)$ into four pieces and then recombining the pieces so as to form regions with greater area than $m(A)$, $m(B)$, $m(C)$, and $m(ABC)$ respectively.  The reader is referred to Ref. \cite{HHM11} for the details.
	\end{enumerate}
\end{enumerate}

\section{Discussion}

In the above results, we have replaced the HRT definition of the holographic entanglement entropy with a new ``maximin surface'' definition.  This makes it much easier to prove theorems, because (like the original Ryu-Takayanagi minimal area surfaces) the maximin surface manifestly obeys certain \emph{global} inequalities comparing it to other surfaces.  We anticipate that there are many other useful results which can also be proven using this technique.

Given the existence of maximin surfaces (with a certain technical stability property (section \ref{stable}) under perturbing $\Sigma$), we have shown that these surfaces are equivalent to the HRT extremal surfaces.  This can be used to prove that HRT surfaces obey strong subadditivity and other desirable properties, and indicates that the HRT proposal is probably correct.  Unlike the proof of static strong subadditivity, the results require the use of the NCC in several places.

These results only follow if one either assumes or proves the existence of the needed maximin surfaces.  We have gone part of the way here by proving that the maximin surface is guaranteed to exist on horizonless spacetimes, and spacetimes like the eternal black hole which have Kasner-type singularities.  Furthermore, Theorem \ref{sta} goes most of the way towards showing that at least one maximin surface is a stable minimal surface under perturbations to $\Sigma$.  Hopefully the remaining gaps can be filled by future work.

The results proven above are really theorems about classical general relativity in asymptotically AdS spacetimes, although the interest of the results arise from the AdS/CFT duality.  This classical limit corresponds to the large $N$, strongly coupled regime of the CFT.  However, one expects that many of the properties should continue to hold away from the classical GR limit.\footnote{An exception might be the monogamy of mutual information (Theorem \ref{mono}), which does not generally hold in quantum information theory, but only in the holographic limit.} In particular, strong subadditivity is a basic property of quantum information theory, and should be equally true for all unitary CFT's regardless of the value of $N$ or the coupling.

\paragraph{String corrections} Weakening the coupling is equivalent to introducing a nonzero string length $l_s$, which produces classical higher-curvature corrections in the action.  This would make the NCC no longer valid.  There are several obstacles to extending the theorems in this case.  For one, we needed the the Raychaudhuri Eq.~(\ref{Ray}) to get focusing results in order to prove the Second Law for causal horizons (and also the closely related fact that null surfaces shot out from extremal surfaces have decreasing area).  These results have not been shown for most higher curvature gravity theories.  

The simplest case which cannot be field-redefined to general relativity is Lovelock gravity.  In this case, one needs to use the Jacobson-Myers entropy functional instead of the area \cite{JM93, dKP11, HMS11, FPS13, dong13}.  However, the Second Law has not been shown for the Jacobson-Myers entropy beyond first order in metric perturbations \cite{fLove}, and even appears to be false nonperturbatively \cite{JM93,lovelock}.

Also, the Jacobson-Myers functional depends on the extrinsic curvature.  In the proof of strong subadditivity, when one reconnects the surfaces $m(AB)$ and $m(BC)$ as in Fig. \ref{SSAfig}, the resulting surfaces have singular extrinsic curvature.  This could also pose problems for generalizing the proof of strong subadditivity.

\paragraph{Semiclassical Corrections}

The other possible deformation from classical general relativity is to consider subleading corrections in $N$.  This corresponds to the semiclassical approximation in the bulk, i.e. quantum field theory in curved spacetime plus small corrections to the background geometry, controlled by the Planck length $l_p$.  In this regime, it is known that the holographic entropy must be modified to include a term proportional to the bulk entanglement entropy \cite{FLM13, SR, BDHM}.  In other words, we are now interested in extremizing the ``generalized entropy'' $S_\mathrm{gen} = A/4 + S_\mathrm{ent}$ of a surface in the bulk.  $S_\mathrm{gen}$ is the same quantity which increases with time when evaluated on slices of a causal horizon, as shown in many settings \cite{10proofs,null}.  This is known as the generalized second law (GSL), and was considered evidence for the holographic principle even before the discovery of AdS/CFT \cite{susskind94}

Ref. \cite{sing} introduced a proof technique in which one uses the GSL in place of the NCC to place restrictions on a semiclassical spacetime manifold.  The types of proof elements used there are similar to those needed here.  For example, the GSL can be used to show that the causal surface lies closer to the boundary than the surface which extremizes $S_\mathrm{gen}$, just as in Theorem \ref{wm}.  It therefore seems likely that e.g. the proof of strong subadditivity can also be extended to the semiclassical regime in this way.  This topic will be explored in future work \cite{NW}.

\subsection*{Acknowledgments}
This research was supported by the Simons Foundation, and NSF grant PHY-1205500.  I am grateful for conversations with Matthew Headrick, Mark Van Raamsdonk, Don Marolf, Veronika Hubeny, Mukund Rangamani, Rob Myers, Netta Engelhardt, Horacio Casini, and Andrea Prudenziati.

\end{document}